\documentstyle[prl,aps,psfig,multicol]{revtex}
\draft
\newcommand{\be}{\begin{equation}}
\newcommand{\ee}{\end{equation}}

\newcommand{\bary}{\begin{eqnarray}}
\newcommand{\eary}{\end{eqnarray}}
\newcommand{\ga}{\Gamma_{\perp}}

\begin{document}
\title{Neutrino propagation in AGN environment}
\author{Sarira Sahu$^*$ and Vishnu M. Bannur}
\address{ 
Institute For Plasma Research,\\
Bhat, Gandhinagar-382 428, India}
\maketitle
\begin{abstract}
\noindent 
Assuming the violation of  equivalence principle (VEP) 
by ultra high energy AGN neutrinos
we study the effect of random magnetic field fluctuation on 
conversion of electron neutrinos to tau anti-neutrinos.\\
\end{abstract}

\begin{multicols}{2}
\section{Introduction}

 Active Galactic Nuclei (AGN) are the powerful sources for the production of
ultra high energy neutrinos in the Universe\cite{ber,eic} and with the present day detectors
these neutrino fluxes can be detected\cite{frich,sobel}.
AGNs are believed to be fueled by the
gravitational energy of matter accreting to a supermassive black hole 
($10^4 M_{\odot}$ to $10^8 M_{\odot}$) at the AGN core, where gravitational 
energy is converted into luminous energy through the acceleration of high 
energy protons\cite{eic,szabo,prot}. These high energy protons looses their energy through $pp$
collision and also through $p\gamma\rightarrow p+e^-+e^+$ and 
$p\gamma\rightarrow N\pi$ processes in the central region.
The pions decay to neutrinos through the $\pi^{\pm}\rightarrow \mu^{\pm}\rightarrow e^{\pm}$
decay chain.
From the pion decay we can expect twice $\nu_{\mu}$ as
$\nu_e$\cite{prot}. A negligible number of $\nu_{\tau}$ are produced in the AGN
environment. The search for such high energy neutrinos by the neutrino
telescopes DUMAND, AMANDA, NESTOR and BAIKAL are undertaken\cite{frich,sobel}.

     Mechanism undergoing neutrino oscillation typically assume that neutrinos
have non degenerate masses and the weak eigenstates are different from the 
mass eigenstates, thereby permitting oscillation between the various 
flavours\cite{pon}. An alternative approach was proposed by Gasperini\cite{gasp} and independently
by Halprin and Leung\cite{hal}. If Einstein's equivalence principle is violated, 
gravity may not universally couple with neutrinos with different flavours. 
Then if this occurs, the weak eigenstates are distinct from the
gravitational one and neutrino oscillations similar to that of vacuum flavour
mixing due to neutrino masses will take place. Thus the violation of equivalence 
principle (VEP) does not require neutrinos to have nonzero masses to oscillate from
one flavour to another\cite{bahcall}.

Using VEP the solution to the solar neutrino problem\cite{mureika} and the
atmospheric neutrino anamoly\cite{halprin} is studied. VEP effect on laboratory
neutrino oscillation experiment is also analysed\cite{utpal}.
The effect of VEP on AGN neutrino has been studied by Minakata and 
Smirnov\cite{mina}.
Magnetic field effect around AGN have important astrophysical consequence.
Neutrinos having magnetic moment or transition magnetic moment (for transition
between different flavours) can flip their helicity in the 
presence of a magnetic field which has a component perpendicular to the direction of
motion of the neutrino. 
The spin-flavour oscillation of high energy neutrinos of AGN origin due to VEP
and large magnetic field are studied previously\cite{piriz,roy,athar}. 
Here we are interested to
study the  spin-flavour transition of $\nu_e\rightarrow\bar{\nu}_{\tau}$
due to random fluctuation in the magnetic field in the AGN
environment in the presence of VEP. We consider both massive and massless
neutrinos here.

\section{Neutrino Propagation}

The evolution equation for propagation of neutrinos in the medium in the
presence of a magnetic field is given by
\be
i{d\over dt}{\pmatrix {\nu_a\cr \nu_x\cr}}
 = {\pmatrix {V-\delta & \mu B_{\perp}(t)\cr \mu B_{\perp}(t) & 0\cr}}
 {\pmatrix {\nu_a\cr \nu_x\cr}}
\label{evu}
\ee
 where $x=s,b$ (sterile, active) and $b = e,\mu$ and $\tau$.
The $\delta$ is given as
$\delta=\cos2\theta~ \Delta m^2/2E$.
$\mu$ corresponds to its diagonal magnetic moment. On the other hand for
Majorana neutrino $\mu$ is the transition magnetic moment.
$V$ is the difference of neutrino interaction potential.
Here we consider neutrino mixing very small, so that $\cos2\theta= 1$.
For considering the fluctuation in the magnetic field, we can write 
$B(t)=B_0 +{\tilde B}(t)$
where ${\tilde B(t)}$ is the fluctuation over the constant background $B_0$.

 In a previous paper, one of the present authors\cite{sahu} has 
derived the average probability equation
for active sterile/active conversion in the presence of a randomly fluctuating
magnetic field (fluctuation in both $B_{\perp}$ and $B_{\parallel}$ 
components). 
The system  consists of magnetic domain 
structure with a size $L_0$ and the magnetic field is uniform and 
constant within each domain. Fields in different domains are randomly aligned.
For the derivation of the probability equation the simple
delta correlation of the magnetic fields in different domains are assumed. 
For the neutrino conversion length greater than the domain size 
$l_{conv} >>L_0$ (where $l_{conv}\sim 1/\Gamma$, 
$\Gamma={\cal P}\Gamma_W$ and $\Gamma_W$ is the weak interaction rate), 
a neutrino will cross many magnetic field domains before it
flips its helicity. Thus the neutrino will experience an average field before 
it flips its helicity.  
The solution for the probability equation is obtained and found
the necessary condition for the positive definiteness of the neutrino
conversion probability ($0\le{\cal P}(t)\le 1$). The condition is given by
(here we consider only the $B_{\perp}$)
\be
4 \large (\mu B_{\perp 0} \large)^2 + (V -\delta)^2 >  {\frac{4}{3}}\ga^2 
\label{ineq}
\ee
where the transverse magnetic field damping parameter $\ga$ is given by
\be
\Gamma_{\perp}={4\over 3}\mu^2B^2_{rms}L_0.
\ee
$B^2_{rms}=<B^2>$ and $L_0$ is the domain size. 
The inequality in eq.(\ref{ineq}) has to satisfy, irrespective of the form
of neutrino potential and the magnetic field.
The effect of random magnetic field on neutrino propagation as well as their conversion
in the early universe hot plasma and in the supernova core is studied\cite{sahu,valle}.
Here we are interested in the propagation of high energy neutrinos in the AGN in the
presence of its own gravitational potential. The effective potential experienced 
by the neutrinos at a distance $r$ from a gravitational source of mass M due to VEP is
($V=V_G$)  
\be
V_G=\frac{1}{2} E~ \Delta f~ \phi(r), 
\ee
where $\phi(r)=GM/r$.
The VEP is parameterised by a dimensionless parameter 
$\Delta f$ and $E$ is the neutrino energy.
The gravitational effect at the neutrino production site is of 
order $\phi(r)\sim 5\times 10^{-3}$ for a $10^8 M_{\odot}$ black hole.
Then for neutrino energy $E=10^{15}$ eV (1 PeV), the gravitational potential for neutrino is
$V_G\sim 2.5\times 10^{12} \Delta f$ eV. 
The matter density in the vicinity of AGN is
found to be $\rho\sim 10-10^4~eV^4$.
The magnitude of the matter potential for neutrino propagating in the
AGN is $G_F\rho/m_p\sim 10^{-29}-10^{-33}~eV$, where $G_F$ is the Fermi coupling constant.
Comparison of $V_G$ with the matter potential shows that for the gravitational
effect to be dominant over the matter effect,
$\Delta f > 10^{-45}$. The optimal sensitivity on VEP can be achieved by
next generation water Cerenkov detector is $\Delta f\sim 10^{-18}-10^{-16}$.
Also it has been shown that, for PeV neutrinos satisfying the resonance 
will give $\Delta f\sim 10^{-28}\Delta m^2/eV^2$. Considering the vacuum
mixing value $\Delta m^2\sim 10^{-10}~ eV^2$
we obtain $\Delta f\sim 10^{-38}$, which corresponds to
$V_G\sim 10^{-26}$ eV and this is again order 
of magnitude larger than the matter
effect. So in our analysis we will neglect the matter effect on 
neutrino propagation and only consider the gravity effect.

In our analysis we will consider both $\Delta m^2\neq 0$  and 
$\Delta m^2=0$ situations. We are interested to consider the process
$\nu_e\rightarrow\bar{\nu}_{\tau}$. This is because the initial fluxes of these
neutrinos in the AGN are maximally asymmetric
($\bar{\nu}_{\tau}/\nu_e\le 10^{-3}$). So any enhancement in this ratio
can signal the spin flavour conversion of $\nu_e$ to $\bar{\nu}_{\tau}$.
Let us consider first $\Delta m^2\neq 0$.
For neutrino propagating in the presence of a constant magnetic 
field $\bf B$, the conversion probability is given by
\be
{\cal P}(r)=\frac{(2\mu B_{\perp})^2}{(2\mu B_{\perp})^2 + V_T^2}
\sin^2\left ([(2\mu B_{\perp})^2 + V_T^2]^{1/2}\frac{r}{2} \right ),
\label{prob}
\ee
where $V_T=V_G-\delta$.
The potential for the massive neutrinos is
\be
V_G-\delta=(2.5\times 10^{12} \Delta f - 
0.5\times 10^{-15} \frac{\Delta m^2}{eV^2}) eV.
\label{hd}
\ee
The resonance condition is obtained when $V_G-\delta=0$ and this gives the
value of $ \Delta f\sim 2\times 10^{-28}\Delta m^2/{eV^2}$.
Let us assume that neutrinos propagating in the AGN medium, will satisfy the
same resonance condition in the presence of a random magnetic field 
fluctuation. Then in that case the inequality in 
eq.(\ref{ineq}) becomes
\be
 \mu B_{\perp 0} > \frac{1}{\sqrt{3}} \ga.
\label{cond}
\ee
Putting 
\be
\ga=2.2\times 10^{-12} \mu_0^2 \left (\frac{B_{rms}}{G}\right )^2
\left (\frac{L_0}{cm}\right ) eV,
\ee
we obtain 
\be
\frac{B_0}{G} > 0.3\times 10^{-3}\mu_0 \left (\frac{B_{rms}}{G}\right )^2
 \left (\frac{L_0}{cm}\right ), 
\ee
where $\mu_0=\mu/\mu_B$.
We take the constant magnetic field $B_0\sim 10^4$ G. For considering 
domain size $L_0$ of order $\sim 1 pc\simeq 3\times 10^{18} cm$ we obtain 
\be
\mu_0 < 10^{-12} \left (\frac{B_{rms}}{G}\right )^{-2}.
\label{mag}
\ee
For considering fluctuation in the magnetic field is 
of order one Gauss, in different domains of size $\sim 1 pc$, 
then we obtain $\mu\sim 10^{-12}\mu_B$. 
This shows that one Gauss fluctuation in magnetic field in $pc$ scale 
might affect the neutrino propagation in the AGN environment. 
We obtain the average conversion probability ${\cal P}=0.5$ for the above 
parameters
\footnote{We use eq.(25) of ref\cite{sahu} to compute the conversion probability 
numerically.}
(magnetic field, domain size and magnetic moment).
Thus half of the $\nu_{e}$ can be converted into $\bar\nu_{\tau}$
when propagating in the fluctuating magnetic field
of the AGN. 

For considering $4 (\mu B_{\perp})^2 >> (V_G-\delta)^2$, in eq.(\ref{prob}) 
the inequality in eq.(\ref{ineq}) will be the same as given in eq.(\ref{cond}).
Also we will get the same constraint on the magnetic moment as shown 
in eq.(\ref{mag}). But in this case constraint on $\Delta f$ will come 
from the inequality $\Delta f << 2\mu B_{0\perp}/\phi(r) E$ (for $\delta\sim 0$). 
Thus for  $B_{0\perp}\sim 10^4 G$
and $\mu_0\sim 10^{-12}$ we obtain $\Delta f << 10^{-31}$.

If the AGN neutrinos satisfy the resonance condition $V_G-\delta=0$ or
for massless neutrinos $V_G << 2\mu B_{\perp 0}$, we obtain the same
constraint on the neutrino magnetic moment $\mu$ and the conversion 
probability is $\sim 0.5$. But the $\Delta f$ is different in both the cases.
In the first case it depends on the $\Delta m^2$ value and in the second
case on the product $\mu B_{\perp 0}$.

The inter galactic magnetic field is $\sim 10^{-6}G$ and this is very small
to reverse the helicity of the neutrino which is traveling $\sim 100 Mpc$
from the AGN to the earth. So once the neutrino flips its helicity in the
AGN fluctuating magnetic field, it will not re-flip in the inter galactic 
magnetic field.
The initial fluxes of high energy neutrinos originating from the AGN are
estimated to have the following ratios: $\nu_e/\nu_{\mu}\simeq 0.5$ and
$\bar\nu_{\tau}/\nu_e\le 10^{-3}$. So if any enhancement of
${\bar\nu_{\tau}}/\nu_e$ is found correlated to the direction of the AGN source
it might be because of the random magnetic field in the AGN and VEP.

For VEP, the weak eigenstates are different from the gravitational one,
so neutrino oscillation occurs due to flavour non-diagonal coupling of 
neutrinos
to gravity. Thus even the massless neutrinos can also oscillate. Let us consider the 
propagation of massless neutrinos in the AGN medium with the constant 
background magnetic field 
$B_{\perp 0}$  very small compared to the random fluctuation $B_{\perp 0} << B_{rms}$. 
In the central region of the AGN, neutrinos are produced because of the
accelerating high energy protons colliding with the dense radiation 
fields ($p\gamma$) and high energy $pp$ collisions. 
According to the spherical accretion model by Kazanas, Protheroe and Ellison\cite{prot},
close to the black hole the accretion flow becomes spherical and  a shock
is formed. The distance from the AGN center to the shock is called as the shock
radius and is of order $10^{14} ~-~10^{20}~ eV^{-1}\cite{piriz,roy}$. 
In this region we assume that the non linear plasma processes
might be the dominating one and magnetic flux lines will be
co-moving  with the turbulent plasma thus creating the randomness 
in the magnetic field. We also assume that in this region the magnetic 
fields will be having domain structures like the solar spots
with intense magnetic field of order $\sim 10^4 G$ and the domain size $L_0$ is
of  order  $ 10^{14}$ to $10^{20}eV^{-1}$, the shock radius. 
Then the inequality in eq.(\ref{ineq}) will give
\be
V_G > \frac{2}{\sqrt{3}}\Gamma_{\perp}
\ee
For the domain size $L_0\sim 10^{14}~eV^{-1}$ and $\Delta f\sim 10^{-31}$ we 
obtain
$\mu < 10^{-12}\mu_B$ and for $L_0\sim 10^{20}~eV^{-1}$ and the same $\Delta f$ as above
we obtain $\mu < 10^{-15}\mu_B$ .

  If we consider that in the AGN environment there is a magnetic field fluctuation 
$B_{rms}\sim 1$ G in the $pc$ scale over
the constant background of order $10^4$ G, then we obtain
$\mu < 10^{-12}\mu_B$ for  normal resonance case ($V_G-\delta=0$) as well as for
$2\mu B_{\perp 0} >> V_G$. For considering
the magnetic field to be random within the shock radius of order $10^{-14} - 10^{20} ~eV$
we obtain $\mu < 10^{-12}\mu_B$ to $\mu < 10^{-15}\mu_B$.
In this case also we obtain ${\cal P}=0.5$

\section{Conclusions}

   We have studied the effect of random fluctuation of magnetic field and VEP on the
PeV neutrinos originating from AGN. We used the perviously found condition for positive
definiteness of the average neutrino conversion probability to constraint the transition
magnetic moment for $\nu_e\rightarrow \bar{\nu}_{\tau}$. We found that random fluctuation
in the AGN magnetic field with VEP will effect the spin-flavour precession of neutrino.
For $\Delta m^2\ne 0$, if we assume that the condition $V_G-\delta=0$ holds, then for 
$B_{rms}\sim 1$G we obtain $\mu < 10^{-12}\mu_B$. Also for considering 
$2\mu B_{\perp} >> V_G$ we obtain the same constraint on $\mu$. 
For considering $B_{rms} >> B_{\perp 0}$ within the sock radius we 
found that for massless neutrinos the positive
definiteness condition on the conversion probability gives $\mu < 10^{-15}\mu_B$ 
for $L_0\sim 10^{20}~eV^{-1}$ and $\mu < 10^{-12}\mu_B$ for  $L_0\sim 10^{14}~eV^{-1}$ 
respectively. Thus the positive  definiteness condition on the conversion probability
of $\nu_e\rightarrow\bar{\nu}_{\tau}$ put constraints on the neutrino magnetic moment.
But this constraint on the magnetic moment is not the physical constraint, because
it comes from the positive definiteness of the conversion probability not from any
physical conditions. But the average conversion probability calculation shows that,
maximum half of the original $\nu_e$ can be converted into ${\bar\nu}_{\tau}$ when
propagating in the magnetic field of the AGN. 
So if any enhancement of
${\bar\nu_{\tau}}/\nu_e$ flux is found correlated to the direction of the AGN source
it might signal the combined effect of VEP and random fluctuation in the magnetic field.
Thus the new generation neutrino telescopes DUMAND, AMANDA, NESTOR and BAIKAL might 
be able to shed more light on this.

\end{multicols}
\end{document}